\documentclass[epjCONF]{svjour}
\usepackage{graphics}
\usepackage[varg]{txfonts} 
\usepackage[latin1]{inputenc}
\session-title{Hot and Cold Baryonic Matter -- HCBM 2010}

\newcommand{\Bene}{Benemerita Universidad Autonoma de Puebla, Puebla, Mexico}
\newcommand{\CERN}{CERN, Geneva, Switzerland}
\newcommand{\Chicago}{Chicago State University, Chicago, IL, USA}
\newcommand{\ELTE}{E\"otv\"os University, Budapest, Hungary}
\newcommand{\MEXICNUNA}{Instituto de Ciencias Nucleares Universidad Nacional Aut\'onoma de M\'exico, Mexico, Mexico}
\newcommand{\MEXIFUNA}{Instituto de F\'isica Universidad Nacional Aut\'onoma de M\'exico, Mexico, Mexico}
\newcommand{\RMKI}{MTA KFKI RMKI, Research Institute for Particle and Nuclear Physics, Budapest, Hungary}
\newcommand{\Pusan}{Pusan National University, Dept. of Physics, Pusan, South Korea}
\newcommand{\Bari}{Universita degli Studi di Bari, Dipartimento Interateneo di Fisica "M. Merlin " \& INFN Sezione di Bari, Bari, Italy}
\newcommand{\Houston}{University of Houston, Houston, USA}
\newcommand{\Texas}{University of Texas at Austin, Austin, USA}
\newcommand{\Yale}{Yale University, New Haven, USA}
\begin{document}
\title{VHMPID: a new detector for the ALICE experiment at LHC}
\author{A. G. Ag\'ocs \inst{1}\and R. Alfaro \inst{2}\and G.G. Barnaf\"oldi \inst{1}\and R. Bellwied \inst{3}\and Gy. Bencze \inst{1}\and D. Ber\'enyi \inst{1}\and L. Boldizs\'ar \inst{1}\and E. Cuautle \inst{4}\and G. De Cataldo \inst{5}\and D. Di Bari \inst{5}\and A. Di Mauro \inst{6}\and I. Dominguez \inst{4}\and E. Fut\'o \inst{1}\and E. Garc\'ia \inst{7}\and G. Hamar \inst{1}\and J. Harris \inst{8}\and A. Harton \inst{7}\and L. Kov\'acs \inst{1}\and P. L\'evai \inst{1}\and Cs. Lipusz \inst{1}\and C. Markert \inst{9}\and P. Martinengo \inst{6}\and M.I. Martinez \inst{10}\and M. Mastromarco \inst{5}\and D. Mayani \inst{4}\and L. Moln\'ar \inst{1}\and E. Nappi \inst{5}\and A. Ortiz \inst{4}\and G. Pai\'c \inst{4}\and C. Pastore \inst{5}\and M.E. Patino \inst{4}\and D. Perini \inst{6}\and D. Perrino \inst{5}\and V. Peskov \inst{4}\and L. Pinsky \inst{3}\and F. Piuz \inst{6}\and S. Pochybov\'a  \inst{1}\and N. Smirnov \inst{8} J.~Song \inst{11}\and A. Timmins \inst{3}\and D. Varga \inst{12}\and A. Vargas \inst{10}\and S. Vergara \inst{10}\and G. Volpe \inst{5}\and J.~Yi \inst{11}\and I.-K.~Yoo \inst{11}}

\institute{\RMKI \and \MEXIFUNA \and \Houston \and \MEXICNUNA \and \Bari \and \CERN \and \Chicago \and \Yale \and \Texas \and  \Bene \and \Pusan \and \ELTE }

\abstract{This article presents the basic idea of VHMPID, an upgrade detector for the ALICE experiment at LHC, CERN. The main goal of this detector is to extend the particle identification capabilities of ALICE to give more insight into the evolution of the hot and dense matter created in Pb-Pb collisions. Starting from the physics motivations and working principles the challenges and current status of development is detailed.} 

\maketitle

\section{Introduction}
\label{intro}
\begin{figure}
\centering
\resizebox{0.85\columnwidth}{!}{%
\includegraphics{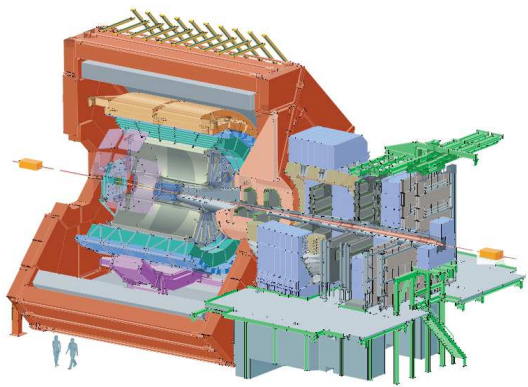} }
\caption{The schematic view of the present ALICE experiment. The VHMPID is likely to take place in the bottom part, around the PHOS detector (shown in purple).}
\label{fig:1}
\end{figure}

The Large Hadron Collider at CERN (Geneva, Switzerland) started its physics program at the end of 2009. It opens the possibility to carry out research in a wide range of topics related to particle physics. One of its large experiments is ALICE (A Large Ion Collider Experiment) \cite{Alessandro:2006yt} (Fig.~\ref{fig:1}.), that is dedicated to study Pb-Pb collisions in order to learn more about the new state of matter characterized by extreme large temperature and energy densities: the quark-gluon plasma (QGP). This knowledge is vital to understand the evolution of the early universe after the Big Bang. One way to gain information about the QGP is to measure jets and high momentum had\-rons. This task requires dedicated detectors. When the presently running ALICE experiment was designed, the new results from the Relativistic Heavy Ion Collider (RHIC, Brook\-ha\-ven, USA) were not yet known, namely that the high momentum had\-rons carry the most important information about the dense and hot matter created in the collisions \cite{star1,phenix1}. In order to extend the particle identification into this region, ALICE needs a new, specific detector, the \textquotedblleft{}Very High Momentum Particle Identification Detector\textquotedblright{} (VHMPID). 
\section{Physics motivation for the VHMPID}

\label{sec:1}
The proposed VHMPID could provide the ALICE experiment with track-by-track charged hadron identity information in the 5 GeV/$c$ $<$ $p_{T}$ $<$ 25 GeV/$c$ region which is not possible with the present ALICE setup. The VHMPID could improve the triggering and tagging of particle showers indicating jet production when combined with other ALICE detectors. This information can be used to study jets in details including their structure (intrajet momentum and flavor correlations \cite{Ellis:1996nv,sickles,Abreu:2000nw,LMolnar:2009}), jet fragmentation fluctuations, parton energy losses and other medium modification effects at unprecedented energies \cite{Barannikova,Ulery}.

These studies are necessary in order to understand in detail the process of hadronization, i.e. hadron matter formation during the QCD phase transition, which is believed to have occurred in the early universe after the Big Bang. Through these measurements it will be possible to follow the evolution of a deconfined quark from its inception in the QGP to its final confinement in an identifiable hadron that will hit the VHMPID. Long sought after experimental verification of QCD principles, such as asymptotic freedom, chiral symmetry restoration and confinement will be achievable. Results in these topics can improve our understanding of the evolution of the universe, the principles of QCD and plasma physics.

\section{The working principle}
\label{sec:2}
\begin{figure}
\centering
\resizebox{0.85\columnwidth}{!}{%
\includegraphics{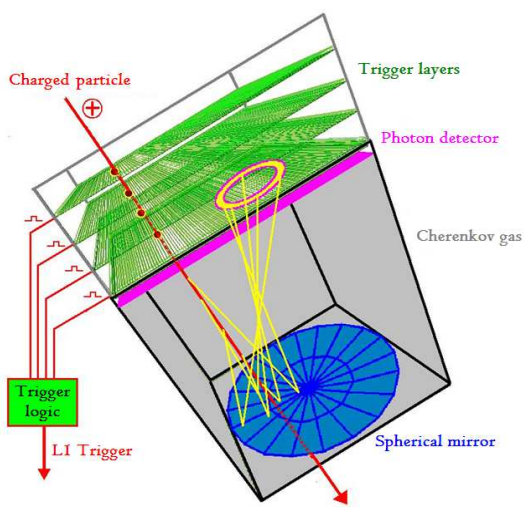} }
\caption{Operation of the VHMPID detector}
\label{fig:2}
\end{figure}

The VHMPID is a Ring Imaging Cherenkov (RICH) detector. Its schematic design is shown in Fig.~\ref{fig:2}. It is designed to identify charged hadrons, mostly protons, pions and kaons. The high-momentum range is achieved by using gas (C$_{4}$F$_{10}$) as the Cherenkov radiator medium. The radiated photons are reflected by a spherical mirror and focused to a ring on the photon detection part. Photon detection is achieved by photosensitive CsI cathoded multiwire proportional chambers or Gas Electron Multiplier (GEM) detectors \cite{Paolo}. These provide the information on the photons in digital form. During the reconstruction, a ring is fitted on the circularily distributed photon hits, from which the Cherenkov angle can be determined. From the angle, the velocity of the particle can be calculated. Combining the velocity with the momentum measured by the tracking detectors, the mass can be calculated, so the incident particle can be identified.

\section{The high-momentum trigger system}
\label{sec:3}

The readout time of the tracking detectors (especially the TPC) is relatively long related to the time between successive collisions provided by the LHC: the expected central collision rate in Pb-Pb is 800 Hz but the TPC can only be operated at 200 Hz \cite{TPC}. Furthermore, data processing and storage constraints are preventing us from keeping all the data from all the collisions, so we are forced to select the interesting events worth storing. For these reasons the VHMPID needs a dedicated high momentum trigger that could be provided by the TRD detector or by a dedicated sub-detector called the High $p_{T}$ Trigger Detector (HPTD). 

It could enhance up to a factor of 40 the recorded high $p_{T}$ events and provide information for particle tracking and impact point reconstruction. The high $p_{T}$ particles are detected by the bending of their path in the 0.5 T magnetic field of ALICE. Because these trajectories just slightly differ from a straight line, the HPTD needs several layers in front of and behind the RICH module in order to detect the bending. In Fig.~\ref{fig:2}. only the frontier layers are shown.

\section{Design questions of the VHMPID}
\label{sec:4}
In order to achieve the high $p_{T}$ identification goals with a Che\-ren\-kov detector, a gasous radiator media is used. In favor of having a reasonable detected photon yield despite the low index of refraction, focusing mirrors with good UV reflective properties are used to map the photons into a ring (in contrast to a mirrorless layout, where the image would be a disk). The resulting image will take place in only a limited area, so only this area is required to be photosensitive. However, manufacturing such large mirrors is difficult and expensive so multiple mirrors or segmented layout should be used. This may increase the required photosensitive area and may cause rings to be splitted into parts that are mapped to different detection regions. This imposes a challenge to the reconstructions algorithms and may reduce the identification performance.

\begin{figure}
\centering
\resizebox{0.85\columnwidth}{!}{%
\includegraphics{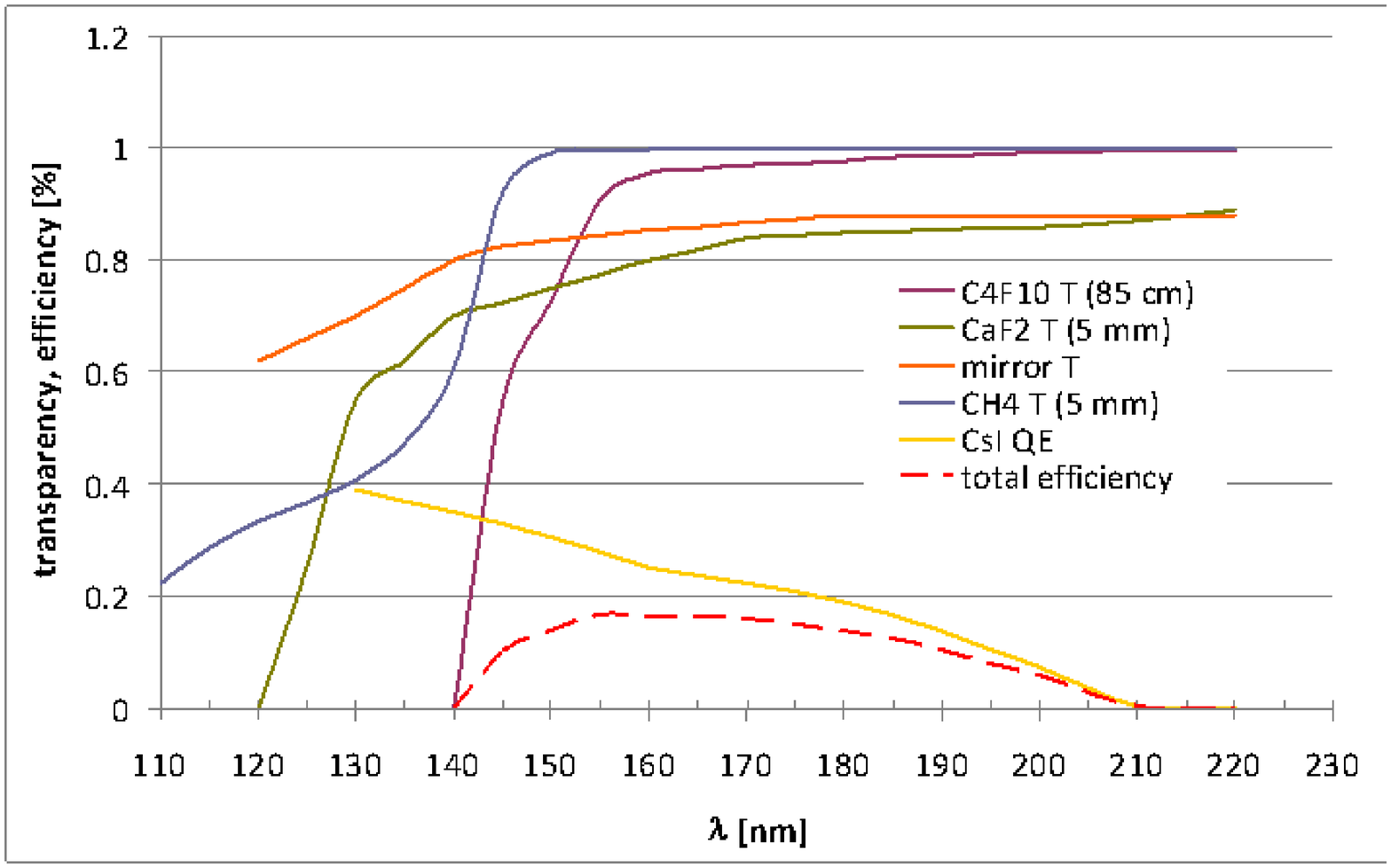} }
\caption{Absorption and efficiency losses in different parts of the detector}
\label{fig:3}
\end{figure}

The total length of the detector is limited by the available space in ALICE. However the performance of both the RICH module and the HPTD is highly dependent on the length. If the length of the radiator media in the RICH module is longer it means that the particle creates more Cherenkov photons on its passage. This also increases the number of photons on the final image, where the ring is more likely to be found by the reconstruction algorithms and the statistical error of the resulting angle will be less. Also, if the HPTD has more length, then the sample points on the trajectory of the particles will more likely to resolve the bending of the track. So the lengths of these parts should be optimized for maximum overall performance.

Also, the various optical media passed by the Che\-ren\-kov radiation must be chosen carefully: absorption can occur in the Cherenkov gas, on the mirror surface and in the transparent window separating the radiation gas and the methane filled photon detection part (Fig.~\ref{fig:3}.). Even if a photon survives these, the quantum efficiency of the CsI might prevent us from detecting it. Since both the absorption and the quantum efficiency is highly wavelength dependent and the CsI is the best candidate for photon detection, the other materials should be chosen such that they minimally absorb the photons in the detection range of CsI.

Absorption is an important issue to keep track of: on average, only a few times ten photons are expected to be detected, but the reconstructed photon number is smaller because of geometrical overlaps (finite resolution of the detector). After all losses, there must be at least three photon hits to fit a ring on them.

\section{Fine tuning the VHMPID with simulations}
\label{sec:5}
\begin{figure}
\centering
\resizebox{0.85\columnwidth}{!}{%
\includegraphics{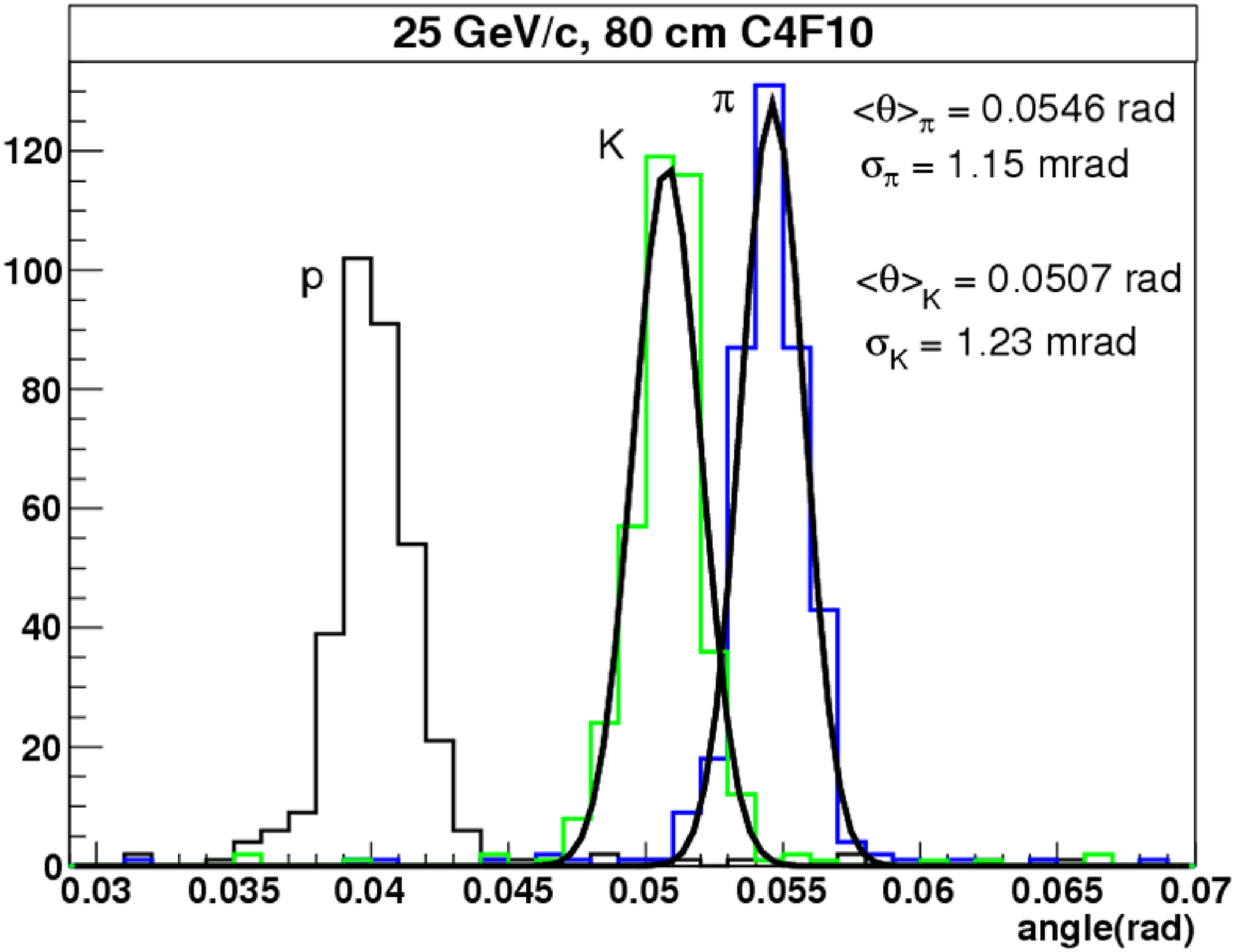} }
\caption{Simulated Cherenkov angles for different particles}
\label{fig:4}
\end{figure}

Building all the possible geometry layouts with all the available materials is impossible and expensive. These scenarios are investigated by detailed computer simulations, so that the best candidates can be selected. The ALICE experiment has a unified system for these tasks: the AliROOT framework \cite{AliROOT}. With its help, the whole VHMPID with all the important materials is constructed in the simulation and thoroughly tested in proton-proton and heavy ion collisions, together with all the other detectors of the experiment. The whole process of the detection is modelled and also the reconstruction of tracks and the particle identities from the measured data. Fig.~\ref{fig:4}. shows the reconstructed Che\-ren\-kov angles for three particle species based on results from the simulations. These reconstructed particle identities are compared to ones that crossed the detector in the simulation and thus the detector efficiency and performance is determined. Based on these, the geometrical and material parameters are tuned to find the optimal solutions.

\section{Prototype test and integration into ALICE}
\label{sec:6}
\begin{figure}
\centering
\resizebox{0.85\columnwidth}{!}{%
\includegraphics{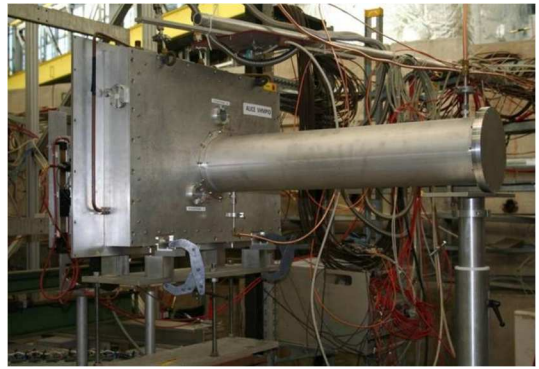} }
\caption{The radiator of the prototype detector. The mirror is located at the end of the tube.}
\label{fig:5}
\end{figure}

\begin{figure}
\centering
\resizebox{0.85\columnwidth}{!}{%
\includegraphics{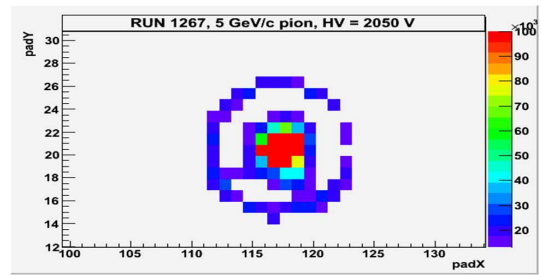} }
\caption{5 GeV/$c$ pion beam crossing the prototype during the beam-test \cite{Antonello}. Place of the incident particle is visible in the middle while the Cherenkov ring is located around it.}
\label{fig:6}
\end{figure}

Theoretical calculations and simulations help to choose the best layouts which can be built in reality to test with real particle beams to measure any differences from the predictions. Also the reconstruction algorithms can be tested on real data. In the fall of 2009 a successful beam test of a prototype RICH module was carried out at CERN (Fig.~\ref{fig:5}.). The earlier investigated \cite{HMPID} CsI coated MWPC type readout was clearly able to see the Cherenkov rings made by the pion beam (Fig.~\ref{fig:6}.).

The latest prototype test was carried out in the fall of 2010 at CERN. For this test a RICH module with dimensions $64 \times 40 \times 80$ cm$^3$ was built implementing the focusing layout. It had two different photon detection sections to test two different window materials. The RICH module was between multiple layers of HPTD chambers. These were used to give tracking information for locating the incident particles. Analysis and data aquisition software was developed to match the RICH ring and particle position to the tracks determined by the HPTD. The data taken at this test is currently being analysed.

\section{Conclusion}
\label{sec:7}
The ALICE experiment at LHC is in need for a new detector to identify very high momentum charged hadrons, so that it can carry out detailed measurements on the most interesting part of some new physics phenomena. This new detector, the VHMPID is being developed by an international collaboration consisting of scientists from CERN, Hungary, Italy, Korea, Mexico and USA. The development is in an advanced stage and already passed several beam tests.

\section*{Acknowledgements}
\label{sec:8}
This work was supported in part by the Mexico project PAAPIT IN115808 and Conacyt P79764-F, the National Research Foundation of Korea (NRF), Hungarian Scientific Research Fund (OTKA) grants NK778816, CK77719, CK77815, and E\"ot\-v\"os University. One of the authors (GGB) is also thanks to J\'anos Bolyai Research Scholarship of the HAS and OTKA PD73596.

\end{document}